\documentclass[sigconf]{acmart}
\renewcommand\footnotetextcopyrightpermission[1]{} 
\usepackage{soul,color}
\usepackage{multirow}
\usepackage{subfig}
\AtBeginDocument{%
  \providecommand\BibTeX{{%
    \normalfont B\kern-0.5em{\scshape i\kern-0.25em b}\kern-0.8em\TeX}}}

\makeatletter
\def\@copyrightspace{\relax}
\makeatother

\setcopyright{none}
\makeatletter
\def\@copyrightspace{\relax}
\makeatother
\begin{document}

\title{PREMIER:  Personalized REcommendation for Medical prescrIptions from Electronic Records}

\author{Suman Bhoi}
\affiliation{\institution{National University of Singapore}}
\email{sumanbhoi@u.nus.edu}

\author{Lee Mong Li}
\affiliation{\institution{National University of Singapore}}
\email{leeml@comp.nus.edu.sg}

\author{Wynne Hsu}
\affiliation{\institution{National University of Singapore}}
\email{whsu@comp.nus.edu.sg}
\begin{abstract}
   The broad adoption of Electronic Health Records (EHR) has led to vast amounts of data being accumulated on a patient's history, diagnosis, prescriptions, and lab tests. Advances in recommender technologies have the potential to utilize this information to help doctors personalize the prescribed medications.
	 In this work, we design a two-stage attention-based personalized medication recommender system called PREMIER which incorporates information from the EHR  to suggest a set of medications. Our system takes into account the interactions among drugs
	 in order to minimize the adverse effects for the patient. 
 We utilize the various attention weights in the system to compute the contributions from the information sources for the recommended medications.
	 Experiment results on MIMIC-III and a proprietary outpatient dataset show that PREMIER outperforms state-of-the-art medication recommendation systems while achieving the best tradeoff between accuracy and drug-drug interaction. Two case studies are also presented demonstrating that the justifications provided by PREMIER are appropriate and aligned to clinical practices.
\end{abstract}


\maketitle
\pagestyle{plain}
\section{Introduction}

Technological advances and digitization have led to the increasing availability of patient information for healthcare data analytics. 
The large repositories of Electronic Health Records (EHR) contain comprehensive information about  patient visits over time, procedures and lab tests results, disease diagnosis, prescribed medications, etc. 
Research on EHR has ranged from
disease inference~\cite{disease_inference3}, outcome prediction~\cite{ma2018kame, hosseini2018heteromed}, to readmission and mortality estimation~\cite{readmission, tan2019ua}.

The White House Precision Medicine Initiative\footnote{https://obamawhitehouse.archives.gov/precision-medicine} has propelled the move to develop and personalize treatment regimens to improve health outcomes.  The traditional ``one size fits all" concept is being superseded by treatments that are tailor-made for each individual. 
One line of research in this direction is the personalization of  medication recommender systems.
In the largest publicly available benchmark dataset MIMIC-III \cite{MIMICIII}, a patient has on average of 20 medications.
Eliminating all the potential drug-to-drug interactions remains a challenge.
While a clinician tries to ensure that the  medications prescribed do not have the common drug-to-drug interactions, 
having a medication recommender system suggest  a list of medicines  given a patient's diagnosis, test procedures, etc that minimizes adverse drug reactions will serve as an important decision support tool to the clinician. 

Besides recommending a list of medications, providing justifications for these recommendations  is  crucial to build trust, thereby increasing  clinicians' acceptance of the system and patients'  adherence to the medications. Hence, an ideal personalized recommender system should
(a) incorporate patient-specific information such as diagnosis, procedures underwent, etc culminated over time to personalize the recommendations;
(b) incorporate drug interaction information to minimize adverse drug interaction; and
 (c) provide justification for the recommendations rendered.

Existing recommender systems such as LEAP use the records of patients' current visit and the drug-to-drug interactions to predict a list of medications~\cite{LEAP, instance_based}. 
These systems do not consider information from  patients' previous visits, leading to lower accuracy.
Further, no justification  is provided for the recommendations.
Other recommender systems such as DMNC ~\cite{DMNC} 
 capture the temporal dependencies in patients' past visits when predicting medications.
However, these systems  do not consider the interactions between drugs which may lead to adverse drug events, particularly in patients with co-morbidity ~\cite{multimorb}.
For example, the usage of anti-inflammatory and anti-rheumatic medications  together with sedatives could result in  diabetic retinopathy~\cite{DDI}. 

The GameNet recommender system~\cite{Gamenet}  employs dynamic memory in conjunction with recurrent neural networks to model temporal dependency in patients' EHR. The authors
use Graph Convolutional Networks (GCN) to model the interactions between  drugs. 
However, this approach does not use a weighted combination of patient's diagnosis, procedures and past medications when retrieving relevant drug interaction information, and is unable to provide justification for the recommendations.

In this work, we propose a recommender system for medications called PREMIER that takes into account the past and current visits of patient records,  drug co-occurrences in the EHR,  as well as the known interactions of drugs, to generate personalized medications with justifications while 
minimizing adverse drug reactions.
 There are two key stages in PREMIER.
 The first stage extracts patient-specific information such as the patient's diagnosis, procedures and his/her previously prescribed medications to create a personalized query vector. 
 The second stage takes as input the query vector to retrieve potential drug interactions from the drug interaction repository, as well as  drug co-occurrences from the EHR of all patients, thus ensuring that the predicted set of medications has minimal drug interactions.

 By utilizing a two-level neural attention model instead of the traditional RNN in the first stage, and a graph attention network  \cite{velickovic2018graph} in the second stage, PREMIER is able to identify the top contributors for each medication recommended, thereby giving the 
justifications for the recommendations.
   Experimental results on the benchmark MIMIC-III EHR and a proprietary outpatient dataset demonstrate that PREMIER outperforms state-of-the-art medication recommender systems
 in terms of accuracy with lower DDI.
Our contributions can be summarized as follows:
\begin{itemize}
\item We propose a two-stage recommender system that combines attention-based recurrent network modeling patients' visit with knowledge graph network modeling drug co-occurrences in the EHR and known drug interactions. 

\item We determine the percentage of contribution from  the various information sources, that is, diagnosis, procedures, previously prescribed medications, to identify the key reasons for recommending a particular medication.

\item We perform extensive  quantitative and qualitative analysis of PREMIER on two datasets. Case studies on two patients indicate that the justifications provided  by PREMIER  are aligned with   clinical practices.

\end{itemize}

\section{Related Work}
\label{section:relatedwork}

Research in medication recommender systems can be broadly classified into rule-based, instance-based and longitudinal.

\smallskip
\noindent \textbf{Rule-based recommendations.}
This approach typically relies  on a collection of rules that capture the knowledge of a medical  expert.
Solt and Tikk~\cite{medication_rule} propose rules to extract medication information from discharge summaries.
Chen \textit{et al.}~\cite{medication_rule2}
derive knowledge patterns from medical information such as disease, symptoms, demographics, measurements, etc of a patient to recommend preferred treatments for chronic heart failures.

The work in \cite{rule_related2} attempts to develop adaptive treatment recommendation for adolescent depression based on hard-coded protocols. Lakkaraju \textit{et al.}~\cite{MDP} use Markov Decision Process (MDP) to learn the mapping between patient characteristics and treatments.
In general, rule-based systems cannot scale or generalize well.
Further, it is difficult to add rules to an already large knowledge base without introducing conflicting rules.

\smallskip
\noindent \textbf{Instance-based recommendations.}
In this approach, only the current visits of  patients are used for medication recommendation. 
The work in ~\cite{LEAP}
formulates the medication recommender problem as a multi-instance multi-label learning task and develops the LEAP algorithm. A recurrent decoder is used to facilitate sequential decision making to model drug-disease relation along with DDI from the EHR records. Once this model is trained on patients' EHR, it predicts medications based on the diagnosis of a patient's current visit. 

Wang \textit{et al.}~\cite{instance_based} jointly embed patient demographics, diagnosis, past medications into a lower dimensional space and  use this embedding for recommendation. They formulate the task of drug recommendation as a link prediction problem taking into consideration a patient's diagnosis and adverse drug reactions. 
The work in ~\cite{instance_reco} also exploits  demographics, diagnosis, and medication information and fuse them with a trilinear method. 
All these approaches ignore patients' diagnosis, etc. from the past visits leading to loss in accuracy and hence, do not personalize well.

\smallskip
\noindent \textbf{Longitudinal recommendations.}
This class of recommender systems aims to  exploit the temporal dependencies among a patient's past visits. 
The work in \cite{sun2016data} clusters the medical records of patients' past visits into cohorts based on their treatment similarities  in order to find patterns in the treatment regimens.

Bajor and Lasko propose to use Recurrent Neural Networks (RNN) to predict medications from a patient's clinical history and billing codes in the EHR~\cite{predict_medication}.  
While their model can predict if a drug is being used by a patient, it is unable to recommend multiple medications for  complex disease conditions. The work in~\cite{Retain} proposes an attention-based predictive model to perform various tasks such as disease and medication prediction.
Le \textit{et al.}~\cite{DMNC} introduce DMNC which infuses the concept of memory augmented neural networks~\cite{memorynetworks} with RNNs to help in the handling of long-range dependencies by incorporating two memory modules 
for medication recommendation. 
All these works ignore drug interactions.

Shang \textit{et al.}  propose a system called GameNet ~\cite{Gamenet} which considers patients' longitudinal visit history and drug interactions. They use dynamic memory and Graph Convolutional networks (GCN) to  personalize the medication recommendations.
However, GCN assumes that the interactions between different drugs have the same weights,
which is not the case in practice. For example, the adverse drug event of long-term or permanent paralysis as a result of using an 
anti-inflammatory medication like Ibuprofen together with an anticoagulant medication like Enoxaparin
is clearly more severe compared to having diarrhea after using   Ibuprofen and a constipation medication like Linaclotide.
In addition,
GCN remains a black-box although attempts have been made to provide explanations via feature visualization \cite{xie2019interpreting}.

\section{Methodology}
\label{subsection:preliminaries}

\begin{figure}[htbp]
	\centering
	\includegraphics[scale = 0.27]{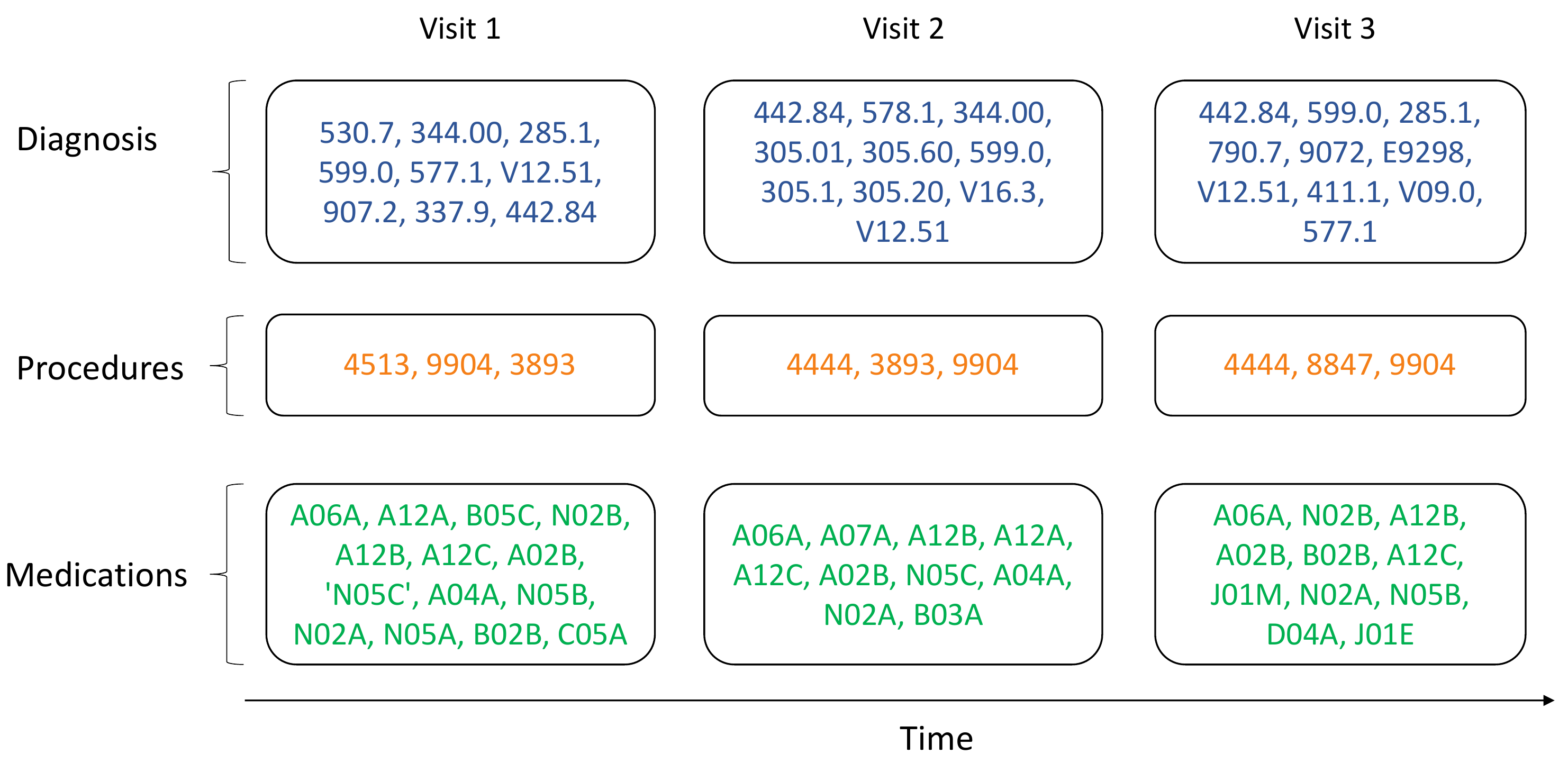}
	\caption{Sample EHR depicting a sequence of three visits.}
	\label{fig:EHR representation}
\end{figure}

The EHR of a patient can be represented as a sequence of $t$ visits.
Figure~{\ref{fig:EHR representation}} shows a sample EHR of a patient with three visits.
The $i^{th}$ visit is given by
 $$\boldsymbol{x}_i = [\boldsymbol{d}_i,\boldsymbol{p}_i, \boldsymbol{m}_i]$$ where  $\boldsymbol{d}_i,\boldsymbol{p}_i, \boldsymbol{m}_i$ are multi-hot vectors capturing the diagnosis, procedure and medication respectively,
$i$ = $1, 2, ...,$ $t$.
Diagnosis and procedures are encoded with ICD-9 system~\cite{slee1978international} while medications are  encoded with ATC classification~\cite{european2013ephmra}.
The total number of diagnosis, procedure and medications are denoted as $N_{d}$, $N_{p}$ and $N_{m}$ respectively.

We use an adjacency matrix  $\boldsymbol{A}_{C}$ to indicate the number of drug co-occurrences in the EHR, in other words, the number of times a pair of drugs have been prescribed together to patients over multiple visits. 
In addition, we also have an 
adjacency matrix $\boldsymbol{A}_{D}$ to indicate whether two drugs are known to have some adverse drug interaction. We construct this matrix from the DDI database~\cite{DDI}. 

Given a patient with 
past visits $[\boldsymbol{x}_1, \boldsymbol{x}_2, ..., \boldsymbol{x}_{t-1}]$, the diagnosis $d_{t}$ and procedure $p_{t}$ codes for the current visit, together with the drug co-occurrence $\boldsymbol{A}_{C}$   and drug interaction information  $\boldsymbol{A}_{D}$, the task is to predict and justify a list of personalized medications for the patient's current visit $\boldsymbol{x}_{t}$.

\begin{figure*}[t!]
	\centering
	\includegraphics[scale = 0.37]{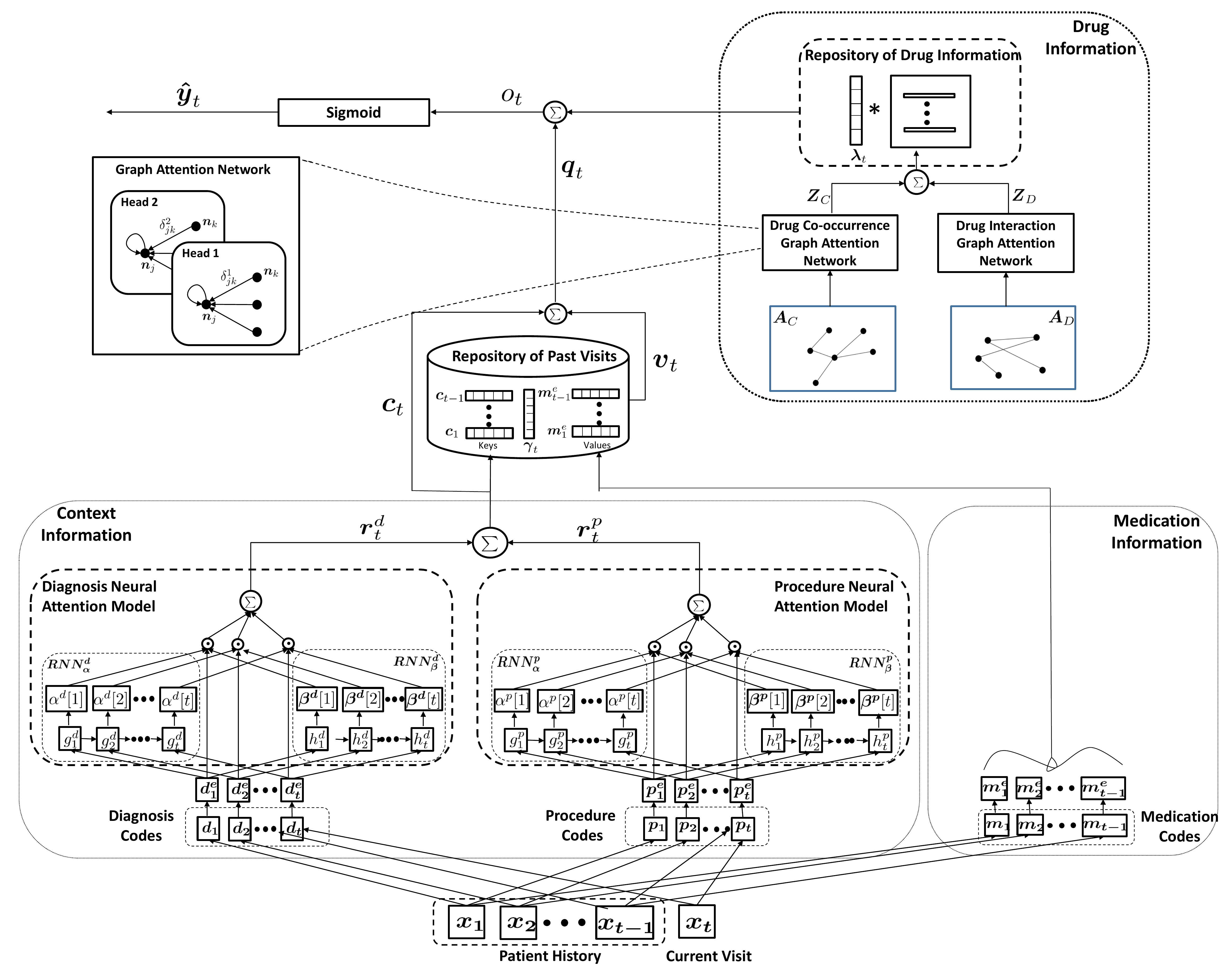}
	\caption{Details of PREMIER.}
	\label{fig:stage1}
\end{figure*}

We observe that a clinician  typically refers to the diagnosis, procedures,  drug-drug interaction, and prescriptions of a patient's past visits (along with diagnosis and procedures of the current visit) when deciding on the appropriate medications for the current visit. Further, this information 
may have different weights depending on their relevance  to the  current visit's diagnosis and procedures.
This motivates us to develop a two-stage recommender system called PREMIER as shown in Figure~\ref{fig:stage1}.

The first stage uses two neural attention models, one for procedures and another for diagnosis.  
This configuration allows us to  alleviate the problem of missing modalities which are common occurrences in EHR. For example, we have patients whose diagnosis codes are available but the procedure codes are absent because the procedures are not recorded or no procedure has been ordered for the patient. 

Each neural attention model has two levels to learn the importance of the various visits, and the importance of the diagnosis (or procedures) of these visits to the current visit. This allows
 varying attention at the visit level as well as at the diagnosis or procedure level within a visit. In other words, our two-level model gives us the flexibility of selecting  which aspect is to be given more attention when predicting medications.
 Further, these attention weights also provide some degree of explanation as opposed to standard RNNs which are hard to interpret.

The second stage  employs graph attention networks 
which allows for the flexibility of incorporating
the frequency of drug co-occurrences and  the severity of drug interactions to learn an efficient representation 
that will improve the accuracy of
the predicted medications while reducing  adverse drug interactions
as well as providing justification for the recommendations.

We describe the details of PREMIER in the following subsections. Table~\ref{table:notations} summarizes the notations used.

\begin{table}[htbp]
	\centering
	\caption{Notations used in PREMIER}
	\label{table:notations}
	\begin{tabular}{|l|l|}
		\hline
		\textbf{Notation}                         & \textbf{Description}   \\ \hline
		$N_{d}$ & Total number of diagnosis\\
		$N_{p}$ & Total number of procedures\\
		$N_{m}$ & Total number of medications\\ \hline
		$\boldsymbol{d}_{i} \in \mathbb{R}^{N_{d}} $  & Multi-hot vector of diagnosis codes  \\ 
		$\boldsymbol{p}_{i} \in \mathbb{R}^{N_{p}}$  & Multi-hot vector of procedure codes   \\ 
		$\boldsymbol{m}_{i} \in \mathbb{R}^{N_{m}}$ & Multi-hot vector of medication codes  \\ \hline 
		$\boldsymbol{A}_{C}\in \mathbb{R}^{N_{m} \times N_{m}}$ & Adjacency matrix of drug co-occurrence\\
		$\boldsymbol{A}_{D}\in \mathbb{R}^{N_{m}\times N_{m}}$ & Adjacency matrix of drug interaction\\\hline
		$\boldsymbol Z_{C} \in\mathbb{R}^{64 \times N_{m}}$                         & Drug co-occurrence representation    \\
		$\boldsymbol Z_{D} \in\mathbb{R}^{64 \times N_{m}}$                        &  Drug interaction representation   \\ \hline
		$ \boldsymbol \alpha \in\mathbb{R}^{t}$                      & Visit level attention     \\ 
		$\boldsymbol \beta\in\mathbb{R}^{64 \times t}$       &  Variable level attention  \\
		$\boldsymbol \gamma_{t} \in\mathbb{R}^{t-1}$       &  Past medication attention \\
		$\boldsymbol \lambda_{t} \in\mathbb{R}^{N_{m}}$       &  Graph network attention \\ \hline
		$\boldsymbol r^d_{t}\in\mathbb{R}^{64}$                       &  Response vector for diagnosis  \\ 
		$\boldsymbol r^p_{t}\in\mathbb{R}^{64}$                       &  Response vector for procedures  \\ 
		$\boldsymbol c_{t}\in\mathbb{R}^{64}$                       &  Context vector   \\ 
		$\boldsymbol v_{t}\in\mathbb{R}^{64}$                      &   Visit history vector  \\ 
		$\boldsymbol q_{t}\in\mathbb{R}^{64}$                       &  Query vector   \\ 
		$\boldsymbol o_{t}\in\mathbb{R}^{64}$                       &  Output vector   \\ \hline
		$ \hat{\boldsymbol y}_{t}\in\mathbb{R}^{N_{m}}$                        & Medication recommendation  \\ \hline
	\end{tabular}
	
\end{table}

\subsection{Create Query Vector}

Given a patient with $t$ visits,  we
linearly embed the diagnosis, procedure and past medication vectors  into a 64-dimensional 
space to learn the distributed representations, denoted as 
$\boldsymbol{ d}^{e}_i$,
$\boldsymbol{ p}^{e}_i$,
$\boldsymbol{ m}^{e}_i$, for $1 \leq i \leq t$ by using the embedding matrices 
$\boldsymbol {E}^{d}$,
$\boldsymbol {E}^{p}$,
$\boldsymbol {E}^{m}$ respectively.
The embeddings of the diagnosis and procedure vectors are passed to the respective neural attention models as shown in  Figure~\ref{fig:stage1}.

Consider a neural attention model similar to~\cite{Retain} that takes as input the diagnosis embedding. We model  the visit-level attention as $\boldsymbol \alpha^d$ where the $i^{th}$ entry, $\alpha^d[i]$, 
is a scalar value depicting the weight for the $i^{th}$ visit, $1 \leq i \leq t$.  
Separately, we also model the diagnostic-level attention of a visit as $\boldsymbol {\beta^d} $ where the $j^{th}$ entry, $\boldsymbol {\beta^d}[j]$,
is a vector depicting the weights of the various diagnostic code embeddings, $1 \leq j \leq t$.

We utilize $RNN^d_{\boldsymbol \alpha}$  to compute the attention vector $\boldsymbol \alpha^d$  as follows: 
\begin{equation}
\label{equation:attention_RNN_alpha}
\begin{array}{l}
\boldsymbol{g}_1, \boldsymbol{g}_2, ..., \boldsymbol{g}_t = RNN^d_{\boldsymbol \alpha}(\boldsymbol{d}^{e}_{1}, \boldsymbol{d}^{e}_{2}, ..., \boldsymbol{d}^{e}_{t}) \\
\boldsymbol \alpha^d = softmax(\phi(\boldsymbol g_{1}), \phi(\boldsymbol g_{2}), ..., \phi(\boldsymbol g_{t}))
\end{array}
\end{equation}
where  each $\boldsymbol{g}_i$ is a hidden layer in $RNN^d_{\boldsymbol \alpha}$ and $\phi$ is a linear transformation  function in which the weight vector $\in \mathbb{R}^{64 \times 1}$.

Similarly, we use a second $RNN^d_{\boldsymbol \beta }$ to  obtain $\boldsymbol {\beta^d}[j]$:
\begin{equation}
\label{equation:attention_RNN_beta}
\begin{array}{l}
\boldsymbol{h}_1, \boldsymbol{h}_{2}, ..., \boldsymbol{h}_{t}  = RNN^d_{\boldsymbol \beta }(\boldsymbol{d}^{e}_{1}, \boldsymbol{d}^{e}_{2}, ..., \boldsymbol{d}^{e}_{t}) \\
\boldsymbol {\beta^d}[j] = \tanh( \psi(\boldsymbol h_{j})) , \quad j = 1, 2, ..., t
\end{array}
\end{equation}
where each $\boldsymbol{h}_j$ is a hidden layer in $RNN^d_{\boldsymbol \beta}$ and $\psi$ is a linear transformation  function in which the weight matrix $\in \mathbb{R}^{64 \times 64}$.

With the generated attentions, we determine the response vector	$\boldsymbol{r}^d_t$ of diagnosis as follows:
\begin{equation}
\label{equation:response_vector}
\boldsymbol{r}^d_t = \sum_{i=1}^{t}\alpha^d[i] \boldsymbol {\beta^d}[i] \odot \boldsymbol d^{e}_{i} 
\end{equation}	
where $\odot$ denotes element-wise multiplication and $\boldsymbol r_{i}  \in \mathbb{R}^{64 \times 1}$.

The response vector for procedure,	$\boldsymbol{r}^p_t$, can be obtained similarly.
With this, we can  define a context vector $\boldsymbol c_t$ as follows: 
	\begin{equation}
\label{equation:context_vector}
\boldsymbol c_t = \boldsymbol r^d_t + w_{1} \boldsymbol r^p_t
\end{equation}
where $\boldsymbol r^d_t$ and  $\boldsymbol r^p_t$ are the response vectors for diagnosis and procedure respectively, and
$w_1$ is a parameter to learn.

We insert  $\boldsymbol c_t$ into a repository of the context vectors for a patient's past visits as well as the corresponding embedded medications which are stored in the form of key-value pairs  where the key is the context vector and the value is the embedded medications.
In other words, the repository consists of  a stack of tuples  $<$$\boldsymbol c_i, \boldsymbol m^e_i $$>$, $i \in [1, t-1]$.

Given the context vector $\boldsymbol c_t$ for the current visit $t$, we compute the attention on the context vectors of the previous visits as follows: 
	\begin{equation}
\boldsymbol \gamma_t = softmax(\boldsymbol c_1\cdot \boldsymbol c_t, ~\boldsymbol c_2\cdot \boldsymbol c_t, ..., \boldsymbol c_{t-1}\cdot \boldsymbol c_t) 
\end{equation}

With this attention, $\boldsymbol \gamma_t  \in \mathbb{R}^{t-1}$, we
obtain the visit history vector $\boldsymbol {v_t}$ where the $i^{th}$ entry is given by	
	\begin{equation}
	\label{equation:history}
	v_t[i]  =   \gamma_t[i] \cdot \boldsymbol m^{e}_{i}       	
	\end{equation}	

Finally, we combine the context vector and the visit history vector to form the query vector $\boldsymbol q_t \in \mathbb{R}^{64} $ as follows:
	\begin{equation}
	\label{equation:query}
	\boldsymbol q_t = \boldsymbol c_t + w_2 \boldsymbol { v_t}	, ~~~w_2\in \mathbb{R}.
	\end{equation}

\subsection{Retrieve Drug Information}
\label{Section:Retrieve_Drug}

Having created the query vector, we will use it to retrieve the drug co-occurrence and drug interaction information.
Recall that we have captured the co-occurrences of drugs in EHR in the form of an adjacency matrix $\boldsymbol A_{C}$, and the known
 drug interaction information in another adjacency matrix $\boldsymbol A_{D}$.

We use  two  graph attention networks, one for drug co-occurrences and another for  drug interactions,
to learn an efficient representation for recommendation (see Figure~\ref{fig:stage1}). This allows PREMIER to model the severity of drug interactions.
Each graph attention network has two layers.
The first layer utilizes two attention heads, while the second layer has one attention head.

We first consider the drug co-occurrence graph attention network. 
The input to a node $\boldsymbol n_j$ in the first layer of this network is the $j^{th}$ row in the feature matrix. Since in this context, the nodes do not have features, an identity matrix $\boldsymbol I$ is used  as the feature matrix depicting the presence of the node itself.
Let $\mathcal{N}_{j}$ be the set of drugs that co-occur
  with $j$ which is found by considering the positions in the $j^{th}$ row of $\boldsymbol A_{C}$ having a non-zero value.  
Then the  output $\boldsymbol n^{e_1}_j$ is given by:
\begin{equation}
	\label{equation:GAT1}
\boldsymbol n^{e_1}_j = \|_{b=1}^{2} \sigma \left( \sum_{k\in\mathcal{N}_{j}}\delta_{jk}^{b} \boldsymbol W^b \boldsymbol n_k \right) 
\end{equation}
where $\|$ denotes concatenation, $\boldsymbol W^{b} \in \mathbb{R}^{64\times N_{m}} $ is the embedding matrix for the $b^{th}$ attention head, 
and $\delta^b_{jk}$ is the attention weight between the pair of drugs $j$ and $k$ computed as follows:

\begin{equation}
	\label{equation:GAT_attention}
\delta^b_{jk} = \frac{ \exp(\text{LeakyReLU} (a^{T} [\boldsymbol W^b \boldsymbol n_j \| \boldsymbol W^b \boldsymbol n_k ]))}{\sum_{f \in \mathcal{N}_{j}} \exp(\text{LeakyReLU} (a^{T} [\boldsymbol W^b \boldsymbol n_j \| \boldsymbol W^b \boldsymbol n_f]))}
\end{equation}

\noindent where $a^T$ is the transpose of the weight vector of a single layer feed-forward neural network.

The final representation of a drug $j$
  is obtained from the second layer of the graph attention network as follows:
 
\begin{equation}
	\label{equation:GAT2}
\boldsymbol n^{e_2}_j = \sigma \left( \sum_{k\in\mathcal{N}_{j}}\delta_{jk} \boldsymbol W \boldsymbol n^{e_1}_k \right)
\end{equation}
where $ \boldsymbol W \in \mathbb{R}^{64\times 128}$ is the embedding matrix. 

With this, we can generate the drug co-occurrence representation $\boldsymbol Z_{C} \in \mathbb{R}^{64\times N_m}$.  In the same manner,  the drug interaction graph attention network is used to  obtain the drug interaction representation $\boldsymbol Z_{D} \in \mathbb{R}^{64\times N_{m}}$.

\smallskip	
We can compute the attention $\boldsymbol \lambda_t$ based on these representations and the query vector $\boldsymbol q_t$  as follows:
	\begin{equation}
	\boldsymbol \lambda_t = softmax( (\boldsymbol Z_{C}  + w_3 \boldsymbol Z_{D} )^{T}\cdot \boldsymbol q_t)
	\end{equation}
	where $(\boldsymbol Z_{C}  + w_3 \boldsymbol Z_{D} )^{T}$ is the transpose of $(\boldsymbol Z_{C}  + w_3 \boldsymbol Z_{D} )$ and $w_3$ is a parameter to learn.

\smallskip	
The final output vector $\boldsymbol o_t$ is obtained as:
	\begin{equation}
	\label{equation:final_output}
	\boldsymbol	o_t = 	\boldsymbol  q_t + w_{4} ( (\boldsymbol Z_{C}  + w_3 \boldsymbol Z_{D} ) \cdot 	\boldsymbol \lambda_t )  , ~~~w_4\in \mathbb{R}
	\end{equation}

\smallskip	
We obtain a vector $\boldsymbol {\hat{y}}_{t}$ by passing $\boldsymbol o_t$ through a linear transformation followed by a sigmoid function.
 A medication $k$ is considered predicted if $\boldsymbol {\hat{y}}_{t}[k] > 0.5$, $k \in \{1, N_{m}\}$.
Using Equations~\eqref{equation:context_vector} and ~\eqref{equation:query}, we have
\begin{equation}
\label{equation: reco_vector}
\boldsymbol {\hat{y}}_{t} =  F \left(\boldsymbol r^d_t + w_{1} \boldsymbol r^p_t + w_2 \boldsymbol v_t	 + w_{4} ( (\boldsymbol Z_{C}  + w_3 \boldsymbol Z_{D} ) \cdot 	\boldsymbol \lambda_t )\right)
\end{equation}

\subsection{Justify Recommended Medications}
\label{Section:contribution_calc}
\enlargethispage{\baselineskip}
In this section, we describe how we justify the recommended medications by determining the percentage of contribution from diagnosis, procedures and previously prescribed medications, and highlighting the top contributors.

Let us consider the  $j^{th}$ medication recommended at the  current $t^{th}$ visit.
Combining the  first term in Equation~\eqref{equation: reco_vector} with
Equation~\eqref{equation:response_vector}, we obtain the contribution due to a patient's diagnosis:
\begin{equation*}
score\_diagnosis = \|_{i=1}^{t} \left( \alpha^{d}[i] \boldsymbol E^{F}[j,:] \left(\boldsymbol {\beta^{d}}[i] \odot \boldsymbol E^{d}[:,k] \right) d_i[k] \right) 
\end{equation*}
where $d_i[k]$  denotes a specific diagnosis in the $i^{th}$ visit where $k \in \{1, N_{d}\}$, $\boldsymbol E^{d}$ denotes the embedding matrix for $\boldsymbol{d}_i$ and $\boldsymbol E^{F}$ is the weight matrix used in the linear transformation function $F$ . 

Similarly, we can compute  the contribution  due to the procedures undergone by the patient, from the second term in Equation~\eqref{equation: reco_vector} and Equation~\eqref{equation:response_vector} as follows:  
\begin{equation*}
score\_procedure = \|_{i=1}^{t}  w_{1} \alpha^{p}[i] \boldsymbol E^{F}[j,:] \left(\boldsymbol {\beta^{p}}[i] \odot \boldsymbol E^{p}[:,k] \right) p_i[k] 
\end{equation*}
where $p_i[k] $ denotes a specific procedure in the $i^{th}$ visit where $k \in \{1, N_{p}\}$ and $\boldsymbol E^{p}$ denotes the embedding matrix for $\boldsymbol{p}_i$. 

In order to determine  the  contribution due to the past prescribed medications,
 we substitute Equation~\eqref{equation:history} to the third term  in Equation~\eqref{equation: reco_vector} as follows:
\begin{equation*}
score\_medication = \|_{i=1}^{t-1} w_{2}  \gamma_{t}[i] \left( \boldsymbol E^{F}[j,:] \cdot \boldsymbol E^{m}[:,k] \right) m_i[k]
\end{equation*}
where $m_i[k]$ denotes the specific medication  prescribed in  the $i^{th}$ visit where $k \in \{1, N_{m}\}$ and $\boldsymbol E^{m}$ denotes the embedding matrix for $\boldsymbol{m}_i$. 

We calculate the contributions due to drug co-occurrences and drug interactions as follows:
\begin{equation*}
score\_occurrence = w_{4} \boldsymbol E^{F}[j,:]  \boldsymbol Z_{C}  \cdot \boldsymbol \lambda_t	 
\end{equation*}
\begin{equation*}
score\_interaction = w_{4} w_{3} \boldsymbol E^{F}[j,:]  \boldsymbol Z_{D} \cdot \boldsymbol \lambda_t 	
\end{equation*}
where $\boldsymbol Z_{C}$ and $\boldsymbol Z_{D}$ are obtained from graph attention networks as described in Section~\ref{Section:Retrieve_Drug}.

The final contributions are obtained by normalizing the five scores such that they add up to 1.
We rank these scores and use the top-2  scores as justification for the recommendations.

\subsection{Loss Functions}

There are three loss functions in PREMIER, namely, binary cross entropy loss, multi-label hinge loss, and drug-drug interaction loss.

The binary entropy loss $\text{L}_{\text{entropy}}$ is defined as:
\begin{equation}
\label{equation:Lbce}
\text{L}_{\textup{entropy}} = -\sum_{i=1}^{t}\sum_{k=1}^{N_m} \left(  y_i[k] \log \hat{y}_i[k] + (1- y_{i}[k])\log (1-{\hat{y}}_{i}[k]) \right) \\
\end{equation} 
where $N_m$ is the total number of medications, $ {\hat y}_i[k]$ and $ y_i[k]$ are the $k^{th}$ predicted and ground truth medication for the $i^{th}$ visit.

Let $U$ be the set of positions where the entry in the ground truth medication vector $\boldsymbol y_{t}$ is 1. 
We define the multi-label hinge loss to ensure that the score of the correctly predicted class is higher than the scores of the other predicted class as follows:
\begin{equation}
\label{equation:Lmulti}
\text{L}_{\text{hinge}} = \sum_{i=1}^{t} \sum_{k=1}^{N_m}\sum_{u \in U} \frac{max(0, 1-({\hat{y}}_{i}[u] - \hat{y}_i[k]))}{|U|}
\end{equation}
where $\hat{y}_{i}[k]$ denotes the $k^{th}$ coordinate of the prediction vector at the $i^{th}$ visit.

The drug-to-drug interaction loss, denoted as $\text{L}_{\text{adverse}}$, penalizes the pairs of drugs that have adverse interactions and is defined as:
\begin{equation}
\label{equation:DDI loss}
\text{L}_{\text{adverse}} = \sum_{i = 1}^{t} \sum_{j=1}^{N_m} \sum_{k=1}^{j-1} (\boldsymbol A_{D} \odot \boldsymbol{\hat{y}}_{i}^\intercal \boldsymbol{\hat{y}}_{i}) [j,k]
\end{equation}

The combined loss is then a weighted sum of the  three losses:
\begin{equation}
\label{equation:Lprediction}
\text{L}_{\text{combine}} = \gamma_1 \text{L}_{\text{entropy}} + \gamma_2 \text{L}_{\text{hinge}} + \gamma_3 \text{L}_{\text{adverse}}
\end{equation}
where $\gamma_1 + \gamma_2  + \gamma_3 = 1$.

\enlargethispage{\baselineskip}
\section{Performance Study}

In this section, we present the results of the experiments we carry out to evaluate the performance of PREMIER for medication recommendations.
We use the largest publicly available MIMIC-III~\cite{MIMICIII} EHR dataset which contains clinical data for 7870 neonates (infants) and 38,597 adults admitted to ICU between 2001 and 2008, and 
 captures attributes such as lab reports, vital signs, medications, etc. This dataset uses the ICD-9 coding system for diagnosis, and the NDC (National Drug Code) system for medications.
 
We pre-process the dataset to filter out those patients that have  only one visit. 
 We create the drug interaction adjacency matrix $\boldsymbol A_{D}$ from the TWOSIDES dataset~\cite{DDI} using the top-40 interactions for each drug because these cover the major severe effects, similar to \cite{Gamenet}.
 Since the TWOSIDES dataset uses the ATC Third Level drug classification system,
   we transform the NDC coding of the medications in MIMIC III to ATC coding  and create the drug co-occurrence adjacency matrix $\boldsymbol A_{C}$.
 Table~\ref{table:data statistics} gives the characteristics of the dataset obtained.

\begin{table}[!h]
	\centering
	\caption{Dataset Characteristics}
	\label{table:data statistics}
	\begin{tabular}{|l|c|}
		\hline
		\textbf{Attribute}                     & \textbf{Statistics} \\ \hline
		Number of patients                          & 6350           \\ \hline
		Number of diagnosis                         & 1958           \\ \hline
		Number of procedures                         & 1430           \\ \hline
		Number of medications                        & 151            \\ \hline
			Average number of visits  per patient                  & 2.36           \\ \hline
		Average number of diagnosis  per visit               & 13.63          \\ \hline
		Average number of procedures  per visit               & 4.53           \\ \hline
		Average number of medications  per visit              & 19.85           \\ \hline
		Maximum number of diagnosis per visit                & 39           \\ \hline
		Maximum number of procedures   per visit              & 32             \\ \hline
		Maximum number of medications per visit               & 55             \\ \hline

	\end{tabular}
\end{table}

\begin{table*}[thbp]
	\centering
	\caption{Results of Comparative Study in MIMIC-III.}
	\label{tab: performance comparison}
	\begin{tabular}{|c|c|c|c|c|c|c|c|c|}
		\hline
	\multirow{2}{*}{Methods} & \multicolumn{3}{c|}{Accuracy} & \begin{tabular}[c]{@{}c@{}}Drug\\ Interaction\end{tabular} & \multicolumn{3}{c|}{Tradeoff} & \multirow{2}{*}{\begin{tabular}[c]{@{}c@{}}Average \#\\Medications\end{tabular}} \\ \cline{2-8}
 & AUC & F1 & Jaccard & DDI & DScore$_{\textup{AUC}}$ & DScore$_{\textup{F1}}$ & DScore$_{\textup{Jac}}$ &  \\ \hline \hline
		Logistic Regression  & 0.6549 & 0.6144 & 0.4545 & 0.0792 & 0.7654 & 0.7370 &   0.6085 & 14.21 \\ \hline
		LEAP & 0.6418 & 0.6064 & 0.4435 & \textbf{0.0665} & 0.7606 & 0.7352  &  0.6013 & 19.01 \\ \hline
		DMNC & 0.6637 & 0.6188 & 0.4450 & 0.0899 & 0.7676 & 0.7366 & 0.5977 & 23.60 \\ \hline
		GameNet  & 0.7472 & 0.6561 & 0.4998 & 0.0790 & 0.8250 & 0.7663 & 0.6479 & 25.06 \\ \hline
		PREMIER  & \textbf{0.7795} & \textbf{0.6812} & \textbf{0.5269} & 0.0750 & \textbf{0.8460} & \textbf{0.7845}  & \textbf{0.6713} & 22.65 \\ \hline
	\end{tabular}%
	
\end{table*}

\begin{table*}[thbp]
	\centering
	\caption{Results of Comparative Study in MIMIC-III for patients with few visits ($< 3$).}
	\label{tab: less than 3 visit performance comparison}
	\begin{tabular}{|c|c|c|c|c|c|c|c|c|}
		\hline
	\multirow{2}{*}{Methods} & \multicolumn{3}{c|}{Accuracy} & \begin{tabular}[c]{@{}c@{}}Drug\\ Interaction\end{tabular} & \multicolumn{3}{c|}{Tradeoff} & \multirow{2}{*}{\begin{tabular}[c]{@{}c@{}}Average \#\\Medications\end{tabular}} \\ \cline{2-8}
 & AUC & F1 & Jaccard & DDI & DScore$_{\textup{AUC}}$ & DScore$_{\textup{F1}}$ & DScore$_{\textup{Jac}}$ &  \\ \hline \hline
		Logistic Regression  & 0.6275 & 0.5933 & 0.4422 & 0.0822 & 0.7453 & 0.7207 &   0.5968 & 14.32 \\ \hline
		LEAP & 0.6329 & 0.5955 & 0.4300 & \textbf{0.0664} & 0.7543 & 0.7271  &  0.5888 & 19.86 \\ \hline
		DMNC & 0.6453 & 0.6002 & 0.4343 & 0.0931 & 0.7540 & 0.7223 & 0.5873 & 21.37 \\ \hline
		GameNet  & 0.7303 & 0.6466 & 0.4853 & 0.0799 & 0.8142 & 0.7594 & 0.6354 & 25.27 \\ \hline
		PREMIER  & \textbf{0.7788} & \textbf{0.6803} & \textbf{0.5264} & 0.0746 & \textbf{0.8457} & \textbf{0.7841}  & \textbf{0.6710} & 22.73 \\ \hline
	\end{tabular}%
	
\end{table*}

We divide the dataset into training, validation and test sets in the ratio of 4:1:1. The hyperparameter values are adjusted on the validation set which resulted in the mixture weights  $\gamma_1=0.79, \gamma_2=0.01$ and $\gamma_3=0.2$. The embedding size for our model is fixed at 64 and training is done using Adam~\cite{kingma2014adam} with a learning rate of 0.0001. The size of the hidden layers for the neural attention models 
 is set to 64.  
  Gated recurrent units (GRU) are used for RNN blocks, since it is efficient and widely used for modeling sequential dependencies. In order to avoid overfitting, we apply a dropout~\cite{srivastava2014dropout} on the input embedding layer with a value of 0.4. 
 The best performing model is chosen  based on the performance on the validation set after 40 epochs. The implementation is done in PyTorch and the models are trained on two NVIDIA Titan RTX GPU. The results  reported are on the test set.

\subsection{Evaluation Metrics}
\label{Section:Evaluation metrics}

We evaluate the accuracy of the predictions with the commonly used metrics  AUC \cite{Gamenet}, F1 \cite{Gamenet} and Jaccard similarity coefficient \cite{LEAP}.
We also use the  DDI  metric proposed in \cite{Gamenet} to 
measure the percentage of pairwise drug interactions in the list of predicted medications.
 A low value for DDI is desirable since it indicates fewer drug interactions.

Since our objective is to maximize the number of correctly predicted medications and minimize the drug interactions, 
 we introduce a metric called DScore  that captures the tradeoff between accuracy and adverse reactions:
 \begin{equation}
 \label{equation:avg BS}
 \textup{DScore} = 	\frac{2 \times \textup{ACC} \times (\textup{1 - DDI})}{\textup{ACC}  + (\textup{1 - DDI})} 
 \end{equation}
where $\textup{ACC} $ denotes an accuracy measure.

Depending on whether we use  AUC, F1 or Jaccard as the accuracy measure, we have three variants of DScore, denoted as  \textup{DScore}$_{\textup{AUC}}$, \textup{DScore}$_{\textup{F1}}$, \textup{DScore}$_{\textup{Jac}} $ respectively.
A high DScore indicates that the set of predicted medications is close to the set of prescribed medications with a low drug-to-drug interaction rate.

\subsection{Comparative Experiments}
We compare PREMIER with several baseline  approaches as well as state-of-art medication recommender systems:

\begin{itemize}
	\item \textbf{Logistic Regression}~\cite{LR}. Here, we use the binary relevance implementation of logistic regression with L2 regularization since our recommendation is multi-label in nature. 

	\smallskip
	\item \textbf{LEAP}~\cite{LEAP}. This is a sequential decision making system which provides recommendations based on the current visit.

	\smallskip	
	\item \textbf{DMNC}~\cite{DMNC}. This is a dual memory neural computer that prescribes a set of medications based on the information of the past visits stored in the external memory components.
	\smallskip	
	\item \textbf{GameNet}~\cite{Gamenet}. This  work models the patient data as a collection of his past visits together with drug interaction information.
\end{itemize}

Table~\ref{tab: performance comparison} shows the results of this experiment. 
We see that PREMIER achieves the best performance in terms of 
accuracy and DScores as it takes into account the past prescribed medications and the varying importance of drug interactions.
In terms of DDI, although LEAP has the best performance,  its accuracy is the worst. This is reflected in LEAP having the lowest DScores.

Note that PREMIER does not completely remove the DDI in its list of recommended medications. This is unavoidable as 
there already exists some degree of DDI in the underlying EHR dataset which is used for training and testing the model. In other words, the medications prescribed by clinicians also contain some drug interactions.
In fact, the DDI in the actual prescribed medications in the test set is  0.0750 which is similar to that attained by PREMIER.
Further,  the average number of actual medications prescribed per  visit in the test set is 20.60 which is close to the average number of medications recommended by PREMIER.\\

\begin{figure}[htbp]
	\centering
	\subfloat[Accuracy]{
		\includegraphics[width=0.4\textwidth, 
		clip=true,trim=280 175 330 160]
		{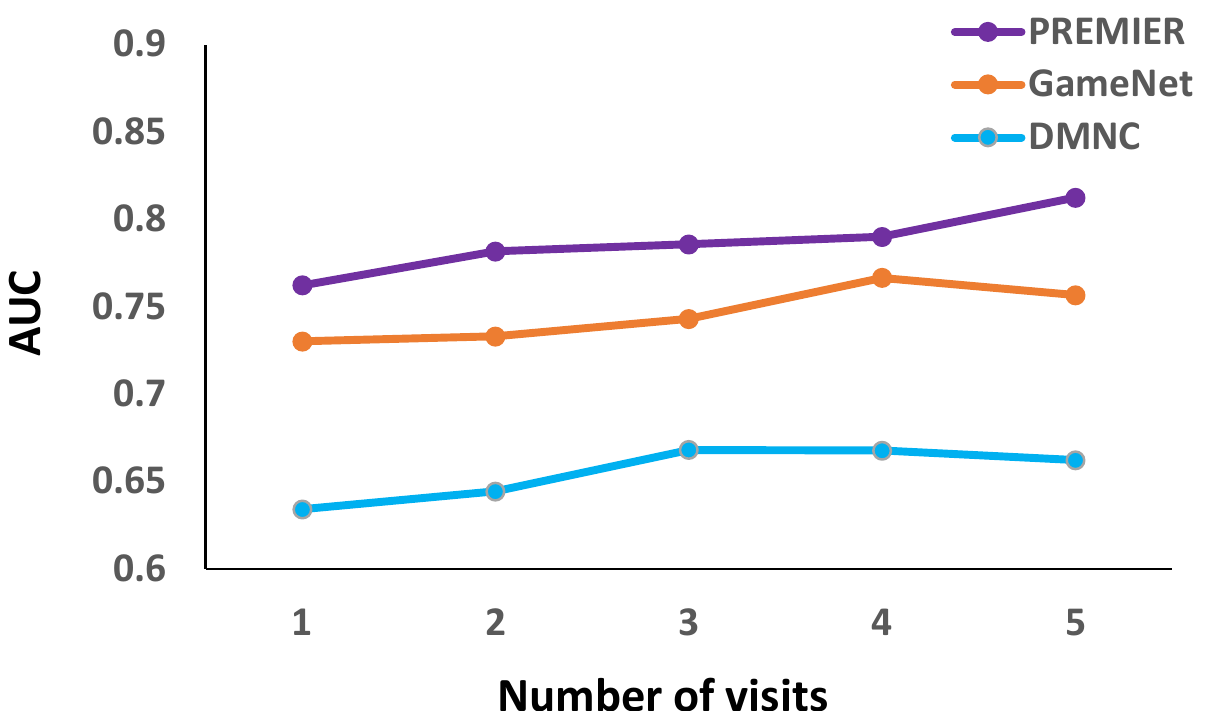}}\\
	\subfloat[Drug Interaction]{
		\includegraphics[width=0.4\textwidth, clip=true,trim=288 170 330 160]{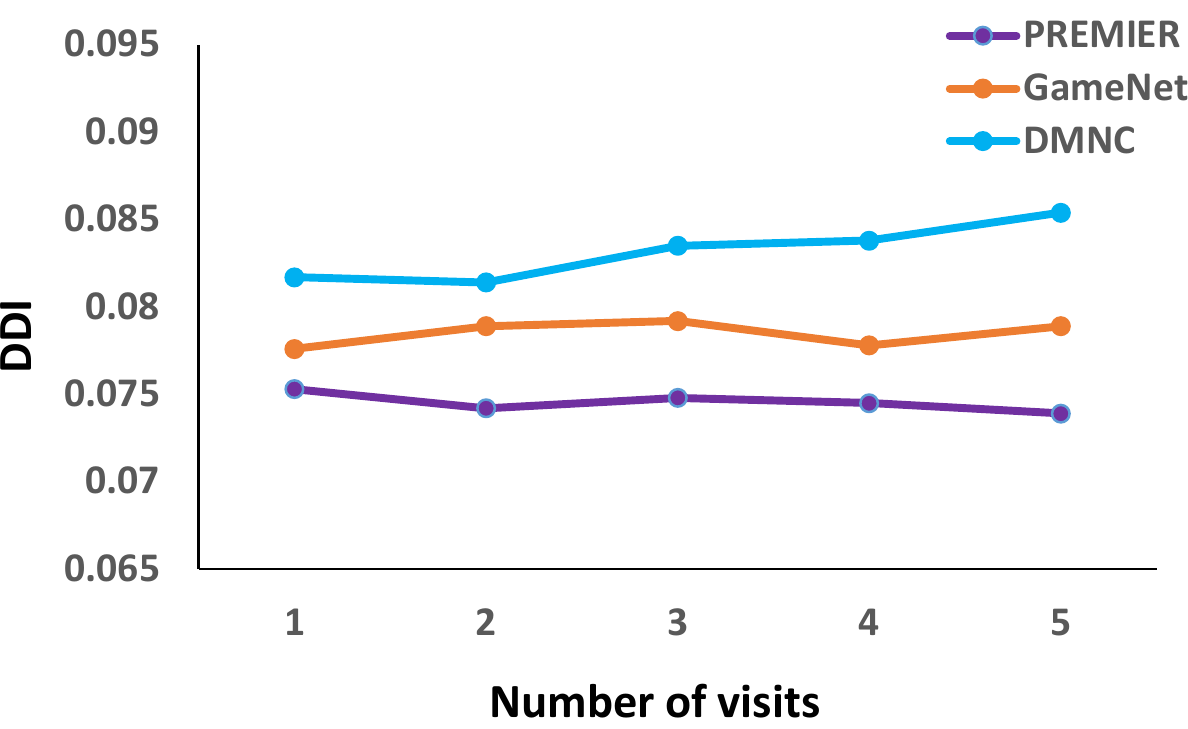}}
	\caption{ Effect of number of visits. }
	\label{fig:visit based analysis}
\end{figure}

\noindent\textbf{Effect of Number of Visits.}
We examine the impact of the number of visits on the performance of various methods.
For this analysis, we use the cohort of patients with exactly 5 visits since there are very few patients with more than 5 visits.
We exclude LEAP and logistic regression from this analysis as they do not use patients' visit information.
Figure~{\ref{fig:visit based analysis}} shows the results.
We observe that PREMIER outperforms GameNet and DMNC in terms of AUC and DDI,
indicating that PREMIER can effectively utilize patients' visit history and strive to recommend safe medications without compromising on the accuracy aspect.

We also compare the performance of the recommender systems for patients with fewer than 3 visits.
 Table~\ref{tab: less than 3 visit performance comparison} shows the results on this cohort of patients. We see that PREMIER continues to be the best performer with almost the same accuracy and DDI, while the other methods have a larger decrease in accuracy and increase in DDI. This indicates that PREMIER remains effective and is robust even for patients with few visits.\\

\noindent\textbf{Number of Recommended Medications.}
Table~\ref{tab:analysis_num} shows the average number of extra medications recommended  as well as the average number of medications that are missed by the various approaches per visit. 
We observe that LEAP misses the most number of medications, suggesting that considering information from previous visits is important.
When we examine the extra  medications recommended, we find the GameNet has the highest number of extra medications, 
indicating that not capturing the severity level of drug interaction may lead to over recommendation.

In contrast, PREMIER has the least number of missed medications per visit. 
The number of extra medications recommended by PREMIER is also relatively low. This suggests that having different attention weights for various instances in multiple information sources while allowing the model to learn the varying severity of drug interactions helps to boost the performance.

 \begin{table}[htbp]
	\centering
	\caption{Average number of missed and extra medications.}
	\label{tab:analysis_num}
	\begin{tabular}{|c|c|c|}
		\hline
		\textbf{Methods}  & \textbf{\# Missed} & \textbf{\# Extra} \\ \hline
		Logistic Regression  & 9.50 & 3.11 \\ \hline
		LEAP  & 8.00 & 7.29 \\ \hline
		DMNC  & 6.96 & 9.22  \\ \hline
		GameNet  & 5.71 & 10.58  \\ \hline
		PREMIER  & 5.49 & 7.55  \\ \hline	
	\end{tabular}
\end{table}

\begin{table*}[thbp]
	\centering
	\caption{Results of Ablation Study}
	\label{tab: abalation study}
	\begin{tabular}{|c|c|c|c|c|c|c|c|}
		\hline
		\multirow{2}{*}{Methods} & \multicolumn{3}{c|}{Accuracy} & \begin{tabular}[c]{@{}c@{}}Drug\\ Interaction\end{tabular} & \multicolumn{3}{c|}{Tradeoff}  \\ \cline{2-8}
		& AUC & FI & Jaccard & DDI & DScore$_{\text{AUC}}$ & DScore$_{\text{F1}}$ & DScore$_{\text{Jac}}$  \\ \hline \hline
		\multicolumn{1}{|l|}{PREMIER {[}diagnosis, procedure{]}} & 0.7456 & 0.6643 & 0.5086 & 0.0810 & 0.8232 & 0.7711 & 0.6548  \\ \hline
		\multicolumn{1}{|l|}{PREMIER {[}diagnosis, procedure, drug repository{]}}  & 0.7473 & 0.6650 & 0.5094 & 0.0781 & 0.8254 & 0.7726 & 0.6562  \\ \hline
		\multicolumn{1}{|l|}{PREMIER {[}diagnosis, procedure, medication{]}} & 0.7508 & 0.6701 & 0.5103 & 0.0790 & 0.8272 & 0.7757 & 0.6567 \\ \hline
		\multicolumn{1}{|l|}{PREMIER} & \textbf{0.7795} & \textbf{0.6812} & \textbf{0.5269} & \textbf{0.0750} & \textbf{0.8460} & \textbf{0.7845}  & \textbf{0.6713}  \\ \hline
	\end{tabular}%

\end{table*}

\subsection{Ablation Study}

In this section, we carry out an  ablation study  to better understand the impact of the various components in PREMIER 
on its  performance. We have the following variants:
\begin{enumerate}
	\item PREMIER[diagnosis, procedure]. This variant uses only the diagnosis and procedure codes of all the visits to predict the list of medications. 
	\smallskip
	\item PREMIER[diagnosis, procedure, drug repository]. This model uses only the diagnosis and procedure codes  along drug information in the repository. It does not use the medications from the past visits of patients.
In other words, patients' previously prescribed medication information is not incorporated in this model.

	\smallskip
	\item PREMIER[diagnosis, procedure, medication]. This model does not have the drug interaction information. In other words, it  uses the diagnosis and procedure codes as well as  the medication prescribed in the past visits of patients.

\end{enumerate}

Table~\ref{tab: abalation study} presents the results  for the different variants of PREMIER. 
As expected, using only diagnosis and procedure information has the lowest accuracy and DScores with the highest DDI.
Incorporating drug co-occurrence and drug interaction information helps to reduce the DDI.
We observe improvements in all the accuracy measures and DScore  when we take into account past medications.

Compared to PREMIER[diagnosis, procedure], we see that the 
 DDI is  lower for other variants because the past medication information implicitly contains some knowledge of drug co-occurrences and drug interactions.
A sharp improvement for both accuracy measures and DDI  is achieved when we consider 
 all the information.

\begin{table*}[h]
	\centering
	\caption{Sample  medications recommended to Patient A and their justifications.}
	\label{tab: med_current_diag1}
	\begin{tabular}{|c|l|l|l|}
		\hline
	&	\textbf{Prescribed Medications} & \textbf{Recommended Medications} & \textbf{Justification} \\ \hline
	1&	Ciprofloxacin & Intestinal  Anti-infectives & \textcolor{blue}{Sepsis (90.36)}, \textcolor{blue}{Septic shock (4.60)} \\ \hline
	2&	Piperacillin and tazobactam &  Beta-Lactam  Anti-bacterials & \textcolor{blue}{Sepsis (39.64)}, \textcolor{orange}{Liver transplant (20.97)} \\ \hline
	3&	Zinc Sulfate & Mineral Supplements & \textcolor{blue}{Sepsis (31.48)},  \textcolor{orange}{Lacrimal gland operated (26.19)}  \\ \hline
4&		Calcium Gluconate & Calcium  & \textcolor{blue}{Septic shock (30.63)}, \textcolor{green}{Omeprazole (20.56)} \\ \hline
5&		Acetaminophen & Analgesics and anti-pyretics & \textcolor{blue}{Delirium (48.52)}, \textcolor{green}{ Opioids (15.36)} \\ \hline

	\end{tabular}
\end{table*}

 \begin{table*}[h]
	\caption{Sample medications recommended to Patient B  and their justifications.}
	\label{tab:med_prev_pro}
	\begin{tabular}{|c|l|l|l|}
		\hline
		&	\textbf{Prescribed Medications}  & \textbf{Recommended Medications} & \textbf{Justification} \\ \hline
		1&		Captopril & Vasodilators  & \textcolor{orange}{Bypass of coronary arteries (71.34)}, \textcolor{blue}{Atrial fibrillation (19.77)}                   \\ \hline
		2&	Acebutolol & Beta blocking agents   & \textcolor{orange}{Bypass of coronary arteries (89.67)}, \textcolor{blue}{Anemia (2.46)}  \\ \hline
		3&	Cisatracurium  Besylate  & Peripherally acting  muscle relaxants & \textcolor{orange}{Mechanical ventilation (90.05)}, \textcolor{blue}{Tracheomalacia} (7.55)                                  \\ \hline
		4&	Diazepam &  Hypnotics and sedatives  & \textcolor{orange}{Mechanical ventilation (77.09)}, \textcolor{orange}{Bypass of coronary arteries (20.00)}  \\ \hline
		5&	Magnesium  Sulfate  & Mineral supplements                                                             & \textcolor{orange}{Mechanical ventilation (57.20)}, \textcolor{blue}{Anemia (30.65)}                                                        \\ \hline
	\end{tabular}
\end{table*}

\begin{table*}[ht!]
	\centering
	\caption{Results on the proprietary test set with an inherent DDI of 0.196, and average number of medications per visit is 4.83.}
	\label{tab: polyclinic performance}
	
	\begin{tabular}{|c|c|c|c|c|c|c|c|c|}
		\hline
		\multirow{2}{*}{Methods} & \multicolumn{3}{c|}{Accuracy} & \begin{tabular}[c]{@{}c@{}}Drug\\ Interaction\end{tabular} & \multicolumn{3}{c|}{Tradeoff} & \multirow{2}{*}{\begin{tabular}[c]{@{}c@{}}Average \#\\Medications\end{tabular}} \\ \cline{2-8}
		& AUC & F1 & Jaccard & DDI & DScore$_{\textup{AUC}}$ & DScore$_{\textup{F1}}$ & DScore$_{\textup{Jac}}$ &  \\ \hline \hline
		Logistic Regression  & 0.5645 & 0.4455 & 0.3291 & 0.3462 & 0.6058 & 0.5299 &   0.4378 &  1.87 \\ \hline
		LEAP & 0.5879 & 0.5176 & 0.3897 & \textbf{0.2600} & 0.6552 & 0.6091  &  0.5105 & 2.89 \\ \hline
		GameNet  & 0.6930 & 0.6092 & 0.4850 & 0.3017 & 0.6956 & 0.6507 & 0.5724 & 7.28 \\ \hline
		PREMIER  & \textbf{0.7189} & \textbf{0.6408} & \textbf{0.5398} & 0.2719 & \textbf{0.7234} & \textbf{0.6816}  & \textbf{0.6199} & 4.27 \\ \hline
	\end{tabular}%
	
\end{table*}

\subsection{Case Study}
\label{Section: case_study}

We present two use cases to demonstrate how PREMIER provides justification for the recommended medications.

\begin{figure}[h]
	\centering
		\includegraphics[width=0.45\textwidth, ]{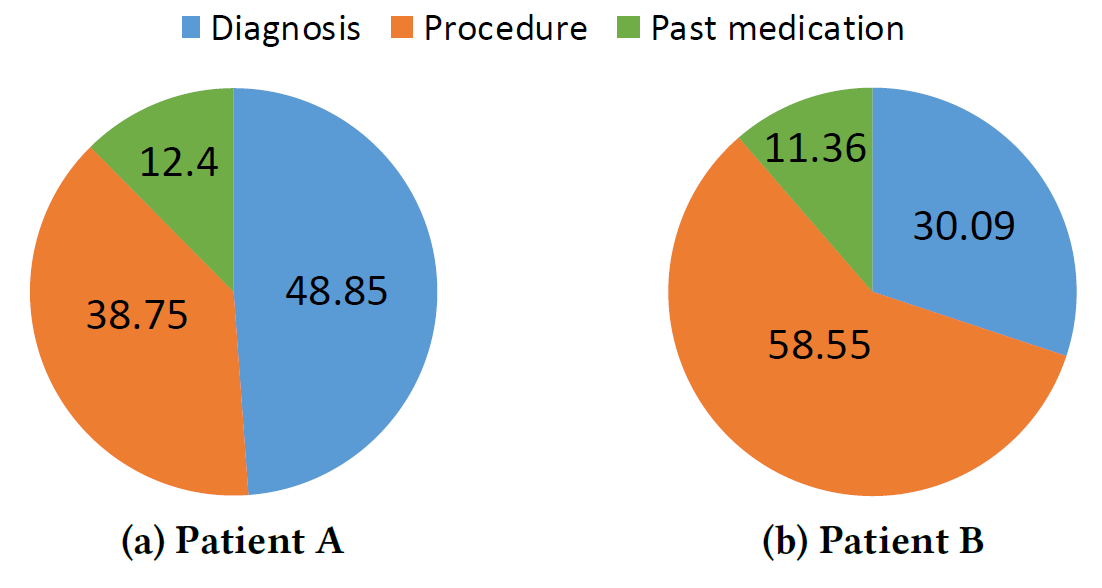}
	\caption{ Distribution of justifications for medications.}
	\label{fig:patients}
\end{figure}

\smallskip
\noindent\textbf{Patient A.}
This patient has been diagnosed with malignant neoplasm of liver and drug-induced delirium, and 
subsequently went on to have a liver transplant surgery. 
The patient returned for a second visit and was diagnosed with septic shock with severe sepsis, a potentially life-threatening condition caused by the body's response to an infection caused by bacteria\footnote{https://www.mayoclinic.org/diseases-conditions/sepsis/symptoms-causes/syc-20351214}. 
 We analyze the medications recommended to Patient A.

Figure~\ref{fig:patients}(a) shows the distribution of the justifications for the  medications recommended to Patient A.
We observe that 
diagnosis contributes  nearly 50\% to the list of medications recommended,  procedures contribute close to 40\% while past medications contribute to the remaining 12\%.

Table~\ref{tab: med_current_diag1} gives a sample of the prescribed medications, recommended medications 
and the corresponding justifications, color-coded as shown in Figure~\ref{fig:patients},  which are 
 based on the top-2 contribution scores obtained from PREMIER.
 We see that the justifications for the
 medications in Rows 1 to 3 are primarily due to the diagnosis sepsis.
 The first two medications Ciprofloxacin and Piperacillin 
 belong to the class intestinal anti-infectives and beta-lactam anti-bacterials respectively.
The justifications make sense as these  medications are known to  treat sepsis~\cite{claessens2007diagnosis}. 

Further, the second highest contributor in Row 2 is the procedure liver transplant which Patient A has undergone previously.
This makes sense as liver transplant is known to increase the risk of sepsis. 
The third medication is  Zinc Sulfate, a  mineral supplement that is often prescribed to prevent  infections or re-occurrence of sepsis~\cite{gariballa2005vitamin}. 
 
Row 4 shows that Calcium is recommended with the justification septic shock, another diagnosis for Patient A. This is reasonable as calcium is typically administered for resuscitation after septic shock~\cite{burchard1992hypocalcemia}.

The medication Acetaminophen in Row 5  is an analgesic and antipyretic, and  the justification shows it is mainly due to the diagnosis delirium.
 Delirium is a serious disturbance in mental abilities that results in confused thinking and reduced awareness of the environment~\footnote{https://www.mayoclinic.org/diseases-conditions/delirium/symptoms-causes/syc-20371386}. 
 Since delirium often stems from pain and fever, analgesics and antipyretics are administered to reduce the pain and body temperature to reduce the severity of delirium~\cite{fosnight2011delirium}.
We see that the  justifications for the medications recommended are appropriate and aligned with clinical practices.

\smallskip \smallskip
\noindent\textbf{Patient B.}
This patient has been diagnosed with coronary atherosclerosis and has undergone a surgery involving the bypass of three coronary arteries. The patient 
is subsequently  diagnosed with severe tracheal disorder and is admitted for a procedure of invasive mechanical ventilation. Figure~\ref{fig:patients}(b) shows that procedure  is the main contributor to the medications recommended. 

Table~\ref{tab:med_prev_pro} shows  the prescribed medications, recommended medications and their corresponding  justifications which is mainly related to the procedure of coronary bypass surgery.
Row 1 is a vasodilator to dilate blood vessels and is used to facilitate the flow of blood
to reduce the pressure on heart after surgery involving coronaries~\cite{kaplan1981peripheral}.
Further, the second highest contributor, atrial fibrillation in Row 1 also makes sense 
as it leads to heart strokes which eventually requires  bypass surgery. 
Row 2 is a  beta-blocking agent to help lower blood pressure, thus  preventing complications after a bypass surgery.

We observe that, a muscle relaxant (Row 3), is recommended by PREMIER and  the justification given is the procedure mechanical ventilation.
This aligns with the fact that muscle relaxants help facilitate the intubation or ventilation procedure~\cite{sharpe1992use}.

The medication in Row 4 is Diazepam which belongs to the class hypnotics and sedatives and is used to  stabilize  patients under ventilation~\cite{grap2012sedation}. 
Since this patient is under invasive mechanical ventilation, mineral supplements (Row 5) are also recommended  to replenish the physiological stability.

The above two cases studies indicate that that PREMIER is able to provide sensible justifications.

\subsection{Experiments on a Proprietary Dataset}  

Finally, we apply PREMIER on a one year outpatient dataset.
Compared to the inpatient MIMIC III, this proprietary  dataset has a larger patient cohort, but the number of diagnosis per patient is much fewer. 
Table~\ref{private} gives the characteristics of this dataset.

\begin{table}[!h]
	\centering
	\caption{ Characteristics of the Proprietary Dataset.}
	\begin{tabular}{|l|c|}
		\hline
		\textbf{Attribute}                     & \textbf{Statistics} \\ \hline
		Number of patients                          & 13640           \\ \hline
		Number of diagnosis                         & 11           \\ \hline
		Number of medications                        & 134            \\ \hline
		Average number of visits  per patient                  & 3.42           \\ \hline
		Average number of diagnosis  per visit               & 2.37          \\ \hline
		Average number of medications  per visit              & 5.04           \\ \hline
		Maximum number of diagnosis per visit                & 7            \\ \hline
		Maximum number of medications per visit               & 8             \\ \hline	
	\end{tabular}
		\label{private}
\end{table}

We divide the dataset into training, validation and testing in the ratio 4:1:1.
The hyperparameter values are adjusted on the validation set which resulted in the mixture weights  $\gamma_1=0.75, \gamma_2=0.05$ and $\gamma_3=0.2$. The embedding size for our model is fixed at 64 with a learning rate of 0.0002 and a dropout value of 0.6. As before, the size of the hidden layers for the neural attention models  is set to 64.  
The best performing model is chosen  based on the performance on the validation set after 30 epochs, and the reported results are for the test set.

Since this dataset does not have procedure codes, we adapt PREMIER and GameNet to run without procedure codes. We omit DMNC as it cannot run without procedure codes. 
 Table~\ref{tab: polyclinic performance} gives the results. We see that PREMIER remains the top performer in terms of accuracy and DScore.
Logistic regression gives the worst results followed by LEAP. However, the DDI for LEAP is the lowest and is closely followed by PREMIER.

In terms of the average number of medications recommended, we observe that LEAP and logistic regression tend to recommend significantly fewer medications to patients while GameNet often over-recommends. On the contrary, the  number of medications recommended by PREMIER is close to that in the ground-truth. 
This gives us the confidence that  PREMIER is robust and can give good recommendations on datasets with different kinds of distributions.

\section{Conclusion and Future Work}

In this work, we have described an end-to-end  system called PREMIER that takes into consideration information from past and current visits along with drug information to make
 accurate recommendations that minimizes adverse drug interactions. The system works in two stages and jointly models the sequential dependencies in past visits and drug co-occurrence and interaction information to derive the recommendations and their justifications.
Experiment results on both a public and private EHR datasets demonstrate the ability of PREMIER to outperform existing solutions in attaining the best tradeoff.
 Case studies also show that the justifications are appropriate and help patients and clinicians understand why the medications are recommended.

Future work includes incorporating lab reports~\cite{labtest} and medical notes~\cite{medicalnotes} which are also a rich source of information and leverage it to gain additional insights such as patient allergies.

\bibliographystyle{ACM-Reference-Format}
\bibliography{bibliography}

\end{document}